\documentclass[aps,twocolumn,pre,floatfix]{revtex4-2}

\usepackage{xcolor,hyperref}
\hypersetup{
   colorlinks,
   linkcolor={blue!50!black},
   citecolor={blue!50!black},
   urlcolor={blue!80!black}
}

\usepackage{epsfig, amsmath}
\usepackage{bm}
\usepackage{soul,ulem}
\usepackage{hyperref}
\usepackage{footmisc}
\usepackage{wrapfig}
\DeclareMathAlphabet{\mathitbf}{OML}{cmm}{b}{it}

\newcommand{\zerovector}{\bm{0}}

\renewcommand{\=}{\!=\!}
\newcommand{\tripleCdot}{:\!\cdot\,}

\newcommand{\ket}[1]{|#1\rangle}
\newcommand{\bra}[1]{\langle #1|}
\newcommand{\braket}[2]{\langle #1|#2\rangle}

\newcommand{\Fv}{\mathitbf F}
\newcommand{\uv}{\mathitbf u}
\newcommand{\xv}{\mathitbf x}

\newcommand{\nv}{\mathitbf n}

\newcommand{\Psiv}{\bm{\Psi}}

\newcommand{\calBold}[1]{\mbox{\boldmath${\cal #1}$}}
\newcommand{\sFrac}[2]{{\textstyle\frac{#1}{#2}}}
\newcommand{\dbar}{{\,\mathchar'26\mkern-12mu d}}

\definecolor{darkGreen}{RGB}{0,100,0}

\begin{document}

\title{Scaling theory of critical strain-stiffening in disordered elastic networks}
\author{Edan Lerner$^1$}
\email{e.lerner@uva.nl}
\author{Eran Bouchbinder$^2$}
\email{eran.bouchbinder@weizmann.ac.il}
\affiliation{$^1$Institute for Theoretical Physics, University of Amsterdam, Science Park 904, 1098 XH Amsterdam, the Netherlands\\
$^2$Chemical and Biological Physics Department, Weizmann Institute of Science, Rehovot 7610001, Israel}

\begin{abstract}
Disordered elastic networks provide a framework for describing a wide variety of physical systems, ranging from amorphous solids, through polymeric fibrous materials to confluent cell tissues. In many cases, such networks feature two widely separated rigidity scales and are nearly floppy, yet they undergo a dramatic stiffening transition when driven to sufficiently large strains. We present a complete scaling theory of the critical strain-stiffened state in terms of the small ratio between the rigidity scales, which is conceptualized in the framework of a singular perturbation theory. The critical state features quartic anharmonicity, from which a set of nonlinear scaling relations is derived. Scaling predictions for the macroscopic elastic modulus beyond the critical state are derived as well, revealing a previously unidentified characteristic strain scale. The predictions are quantitatively compared to a broad range of available numerical data on biopolymer network models and future research questions are discussed. 
\end{abstract}

\maketitle

\section{Introduction}

Disordered elastic networks are composed of a spatially-disordered set of nodes, which are embedded in a $\dbar$-dimensional space and interact through elastic bonds with some prescribed connectivity. Such networks provide a framework for describing a broad range of physical phenomena and systems. These include amorphous materials such as non-Brownian suspensions~\cite{asm_pnas_2012}, proteins~\cite{liu_allostery,WODAK2019566}, polymeric fibrous materials~\cite{RevModPhys.86.995}, athermal biopolymer fibrous networks, such as collagen and fibrin that play crucial roles in various biological contexts~\cite{kees_nature_2005,Janmey_soft_matter_2007,RevModPhys.86.995}, and even confluent cell tissues~\cite{merkel_pnas_2019,Kim2021}. Consequently, understanding their generic physical properties --- including under external driving forces --- is of prime importance. 

Many disordered elastic networks share two generic features. First, the network's degree of connectivity --- i.e., the average number of bonds per node --- is small such that the network is sub-isostatic~\cite{robbie_nature_physics_2016}. That is, such networks are underconstrained, implying that in the absence of additional stabilizing interactions, the network would possess \emph{floppy} (\emph{zero}) \emph{modes}, which are nontrivial, collective deformation modes that do not involve any energetic cost~\cite{maxwell_1864,phonon_gap_2012}. Second, many disordered elastic networks do include additional weak/soft interactions that endow them with finite, yet typically small, elastic response coefficients (constants). That is, these networks typically feature two widely separated rigidity scales. For example, in fibrous networks, stiff (strong) interactions emerge from fiber stretching/compression, while the soft (weak) interactions emerge from fiber bending. 

\begin{figure*}[ht!]
  \includegraphics[width = 1\textwidth]{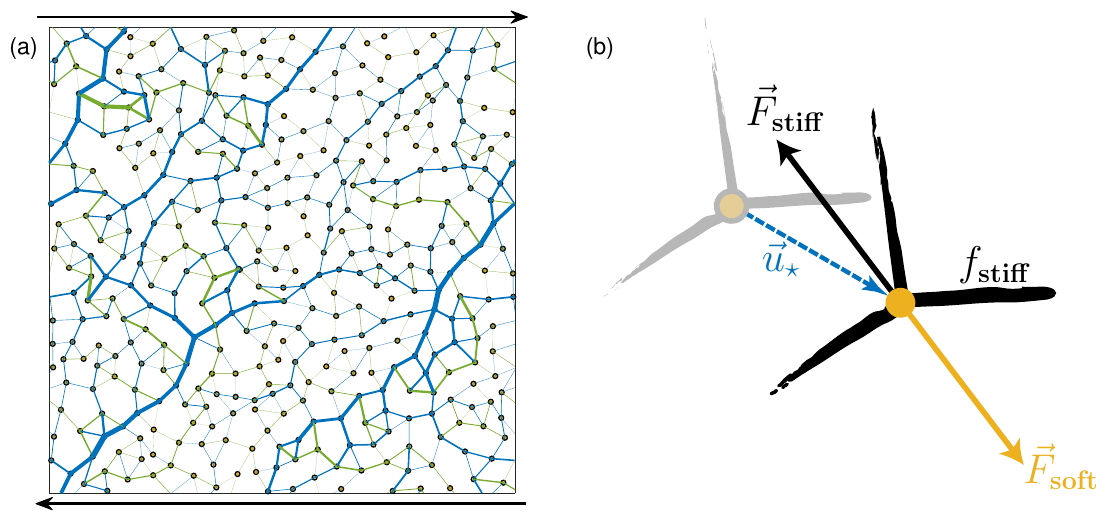}
  \caption{\footnotesize (a) An example of a sub-isostatic disordered elastic network in 2D, featuring a degree of connectivity of $z\!=\!3.8$ bonds per node, after being deformed to the critical strain-stiffened state. The outer arrows denote the externally applied shear deformation. The thickness of the bonds represents components of the state-of-self-stress that emerges at the critical strain-stiffening point, which endows the initial floppy network with a finite shear modulus~\cite{matthieu_thesis}. Blue (green) bonds correspond to stretching (compression) forces. (b) An illustration of the force balance  between the stiff and soft sub-networks at a node. In the absence of the soft sub-network, the representative node in the stiff sub-network is located at the grey circle and the stiff bonds connected to it are shown in grey as well. The introduction of the soft sub-network perturbs the stiff sub-network, resulting in a nodal displacement $\vec{\uv}_\star$ (dashed blue arrow). At the new node's position, the resultant stiff sub-network force $\vec{\bm F}_{\rm stiff}$ (black, obtained from the vectorial sum of the individual bond forces, each of magnitude $f_{\rm stiff}$) is balanced by the resultant soft sub-network force $\vec{\bm F}_{\rm soft}$ (yellow).}
  \label{fig:fig1}
\end{figure*}

These two generic properties, corresponding to the network's topology and its characteristic rigidity scales, are quantified by the average connectivity (bonds per node) $z\!<\!z_{\rm c}$, where $z_{\rm c}\!=\!2\dbar$ is the Maxwell rigidity criterion~\cite{maxwell_1864}, and by the (dimensionless) rigidity scales ratio $\kappa\!\ll\!1$. A remarkable property of disordered elastic networks with these generic properties is that they undergo a dramatic macroscopic stiffening transition when subjected to external strains. In particular, an initially undeformed network that features a vanishing shear modulus $G\!=0$ for $\kappa\=0$ and $z\!<\!z_{\rm c}$, undergoes a sharp stiffening transition as its shear strain $\gamma$ attains a critical value $\gamma_{\rm c}$, upon which $G$ features a jump discontinuity, attaining a finite value $G(\gamma_{\rm c},\kappa=0)$~\cite{wouter_pre_2017,robbie_thesis,merkel_pnas_2019}. This driven phase transition is associated with the emergence of a special internal state, commonly termed a state-of-self-stress (SSS, cf.~Fig.~\ref{fig:fig1}a), which corresponds to a set of putative bond forces that exactly balance each other on the network's nodes~\cite{gustavo_pre_2014,robbie_pre_2018,merkel_pnas_2019}.

Substantial experimental, computational and theoretical effort has been made in order to characterize and understand the strain-stiffening transition in disordered elastic networks~\cite{kees_nature_2005,Janmey_soft_matter_2007,RevModPhys.86.995,gustavo_pre_2014,robbie_pre_2018,merkel_pnas_2019,Truskinovsky_pre_2019,mackintosh_pre_2016,Rens_JPCB_2016,robbie_nature_physics_2016,wouter_pre_2017,robbie_pre_2018,robbie_thesis,maha_prl_2008,mackintosh_prl_2019,fred_arXiv_2022,fred3D_pre_2021,fred_emt2023}. Yet, a fundamental theoretical understanding of this driven phase transition is currently lacking. Here, we first develop a complete scaling theory of the critical strain-stiffened state, and then derive scaling relations that go beyond the critical state. 

The theory is based on conceptualizing the entire network as composed of two interacting sub-networks; a stiff underconstrained sub-network at its sharply defined critical state $\gamma\=\gamma_{\rm c}$ and a soft sub-network of a characteristic dimensionless stiffness $\kappa\!\ll\!1$, where the latter is treated as a perturbation on top of the former, see Fig.~\ref{fig:fig1}b. This framework allows to derive a set of nonlinear scaling relations for all basic quantities that characterize the critical state. We then go beyond the critical state, deriving scaling predictions for $G(\gamma,\kappa)$ away from $\gamma_{\rm c}$, which reveal a previously unidentified characteristic strain scale in disordered elastic networks.

Our theoretical predictions are quantitatively compared to extensive numerical datasets available in the literature, obtained from computer simulations of  a wide variety of biopolymer networks models in 2D ($\dbar\=2$) and 3D ($\dbar\=3$). The comparison reveals excellent agreement with the numerical data, indicating that our scaling theory is relevant for athermal biopolymer networks. It also raises questions concerning the range of critical scaling, in terms of the smallness of $\kappa$ and its dependence on the degree of connectivity $z$. This and other relevant open questions are discussed, along with future research directions.

\section{Results}

\subsection{Scaling theory of the critical state}

Consider first the stiff, underconstrained sub-network in the absence of strain. Being sub-isostatic, this sub-network features floppy (zero) modes, i.e.~modes that can be deformed without any energetic cost. As a shear strain $\gamma$ is applied at a given direction, the system self-organizes anisotropically to give rise to the SSS at $\gamma\=\gamma_{\rm c}$~\cite{gustavo_pre_2014}, cf.~Fig.~\ref{fig:fig1}a. There exist, however, an extensive number of zero modes that are not coupled to the applied strain and remain floppy at $\gamma\=\gamma_{\rm c}$. In more formal terms, one can show that floppy modes are orthogonal to the applied shear strain~\cite{footnote}. Consider then adding the soft sub-network of dimensionless rigidity $\kappa\!\ll\!1$ to the stiff one, where the two sub-networks interact at the network's nodes, see illustration in Fig.~\ref{fig:fig1}b. The soft sub-network may be regarded as a perturbation applied to the stiff one, introducing forces of order $\kappa$. The latter implies that the soft sub-network features a characteristic deformation that is independent of $\kappa$ (but depends on the average connectivity $z$)~\cite{robbie_pre_2018}. As long as the soft interactions --- of characteristic stiffness $\kappa$ --- are random, their precise functional form and physical origin (e.g., angular, bending interactions or random radial interactions) does not effect the scaling laws derived here, as also explicitly demonstrated in~\cite{robbie_pre_2018}.

How would the stiff sub-network respond to the forces introduced to it by the soft sub-network? To address this question, we first note that --- independent of their functional form --- the soft interactions are random in nature, and hence will inevitably have projections on the floppy modes of the stiff sub-network, those that persist at $\gamma_{\rm c}$ as they are not coupled to the strain. As the floppy modes are energetically cheap to move, one expects them to control the response of the stiff sub-network, for example the node displacements (relative to the node positions prior to the addition of the soft sub-network) of characteristic magnitude $u_*$, see Fig.~\ref{fig:fig1}b.

As the response of the stiff sub-network to the random perturbations introduced by the soft sub-network is expected to be dominated by its floppy modes, its energy $U_{\mbox{\tiny stiff}}(u)$ as function of the node displacement $u$ features vanishing first (mechanical equilibrium) and second (floppy normal modes) derivatives at $u\=0$. Since $u\=0$ is a stable state, a cubic term $\sim u^3$ is also excluded. Hence, we expect $U_{\mbox{\tiny stiff}}(u)\!\sim\!u^4$, i.e.~that the energy of the stiff sub-network is dominated by quartic anharmonicity, in line with the arguments recently spelled out in~\cite{manning_rigidity1_pre_2022, manning_rigidity2_pre_2022}. Similar observations in gently compressed ellipses~\cite{elipse_packings_prl_2009} and other nonspherical particle packings~\cite{ohern_pre_quartic_modes_2018}, as well as in deformable particle packings~\cite{ohern_deformable_particles_2021} are available. Consequently, the net force applied to the network's nodes by the stiff sub-network follows $F_{\mbox{\tiny stiff}}\!\sim\!dU_{\mbox{\tiny stiff}}/du\!\sim\!u^3$. As this force balances the imposed soft forces $F_{\mbox{\tiny soft}}\!\sim\!\kappa$, i.e.~$F_{\mbox{\tiny stiff}}\=F_{\mbox{\tiny soft}}$ (cf.~Fig.~\ref{fig:fig1}d), we predict the network to feature node displacements of characteristic magnitude $u_*$ that scales as
\begin{equation}
\label{eq:displacement}
u_*(\kappa) \sim \kappa^{1/3} \ .
\end{equation}

The basic result in Eq.~\eqref{eq:displacement} has immediate implications for the energetics of the system. First, it implies that the energy of the stiff sub-network scales as $U_{\mbox{\tiny stiff}}\!\sim\!u_*^4\!\sim\!\kappa^{4/3}$. Since the soft interaction is characterized by the (dimensionless) rigidity $\kappa\!\ll\!1$, its energy is expected to scales as $U_{\mbox{\tiny soft}}\!\sim\!\kappa$. Taken together, the total energy $U$ of the systems is predicted to scale as
\begin{equation}
\label{eq:energy}
   U(\kappa) = U_{\mbox{\tiny stiff}}(\kappa) + U_{\mbox{\tiny soft}}(\kappa) \sim U_{\mbox{\tiny soft}}(\kappa) \sim \kappa \ ,
\end{equation}
since $U_{\mbox{\tiny stiff}}\!\ll\!U_{\mbox{\tiny soft}}$ in the limit of small $\kappa$. That is, we predict that the total energy of the network is dominated by the energy of the perturbing soft sub-network. 

It is natural to consider next the forces in the problem. To that aim, we distinguish between two characteristic forces. The first one corresponds to the net force of magnitude $F$ applied to each node in the network by the two sub-networks (cf.~Fig.~\ref{fig:fig1}b), $F\=F_{\mbox{\tiny stiff}}\=F_{\mbox{\tiny soft}}$, where the latter equality follows from mechanical equilibrium, as already invoked above (the {\em total} net force at each node obviously vanishes, $F_{\mbox{\tiny total}}\=F_{\mbox{\tiny stiff}}-F_{\mbox{\tiny soft}}\=0$). The second one corresponds to the magnitude of the bond scale force $f$ (cf.~Fig.~\ref{fig:fig1}b), which includes a contribution from the stiff sub-network $f_{\mbox{\tiny stiff}}$ (e.g., corresponding to fiber stretching/compression in fibrous networks) and a contribution from the soft sub-network $f_{\mbox{\tiny soft}}$ (e.g., corresponding to fiber bending in fibrous networks). Obviously, the vectorial sum of the individual stiff bond forces equals the net force applied by the stiff sub-network on the network's nodes and likewise the vectorial sum of the individual soft forces equals the corresponding net force applied by the soft sub-network. 

With this distinction in mind, we set out to derive a scaling estimate for the eigenvalue $\lambda$ of the geometrical operator defining the SSS~\cite{sss_epje_2018}, defined and explained in Appendix~\ref{sec:appendix_sss_operator}. At the critical $\gamma_{\rm c}$ state in the absence of a soft sub-network perturbation, $\kappa\=0$, the geometrical operator associated with the SSS features an identically vanishing eigenvalue, $\lambda\=0$. That means that the eigenvector corresponding to $\lambda\=0$ is interpreted as composed of a set of putative bond forces that exactly balance on the network's nodes. With the application of a soft sub-network perturbation, $\kappa\!>\!0$, the network node displacements $u_*$ predicted in Eq.~\eqref{eq:displacement} emerge and one expects $\lambda$ to become finite as the $\kappa\=0$ state is distorted. Previous work has shown that $\lambda$ can be expressed in terms of the ratio $F/f$ as $\lambda \!\sim\!(F/f)^2$~\cite{asm_pnas_2012}. Furthermore, in Appendix~\ref{sec:appendix_sss_variation} we show that $\lambda$ increases with the node displacements $u_*$ as $\lambda\!\sim (F/f)^2\!\sim u_*^2$. Using Eq.~\eqref{eq:displacement}, we then predict 
\begin{equation}
\label{eq:SSS}
\lambda(\kappa) \sim \kappa^{2/3} \ .
\end{equation}

The above results can be readily used to obtain a prediction for the bond forces $f$, and consequently for the stress $\sigma$ in the network, which satisfies $\sigma\!\sim\!f$. First, recall that $F\=F_{\mbox{\tiny stiff}}\=F_{\mbox{\tiny soft}}\!\sim\!\kappa$ and that $f_{\mbox{\tiny soft}}\!\sim\!\kappa$. Consequently, Eq.~\eqref{eq:SSS} --- i.e.~$\lambda\!\sim\!(F/f)^2\!\sim\! \kappa^{2/3}$ --- can be satisfied if $f$ is dominated by the stiff bond forces, i.e.~$f\!\sim\!f_{\mbox{\tiny stiff}}\!\ll\!F_{\mbox{\tiny stiff}}$. Therefore, we obtain $f_{\mbox{\tiny stiff}} \sim \kappa/\sqrt{\lambda}$, which implies that the network's stress $\sigma$ satisfies   
\begin{equation}
\label{eq:stress}
\sigma(\kappa) \sim f_{\mbox{\tiny stiff}}(\kappa) \sim \kappa^{2/3}\qquad\hbox{with}\qquad f_{\mbox{\tiny stiff}}\gg f_{\mbox{\tiny soft}} \ .
\end{equation}

Quite counter-intuitively and surprisingly, the last relation in Eq.~\eqref{eq:stress} implies that the while the overall network's energy $U$ is dominated by the soft sub-network, cf.~Eq.~\eqref{eq:energy}, the overall network's stress $\sigma$ is dominated by the stiff sub-network. This intriguing result is a manifestation of the singular nature of the response of the critical strain-stiffened state to the perturbation introduced by the soft sub-network, to be further discussed below. 

The most pronounced macroscopic effect of the strain-stiffening transition is the dramatic increase in the shear modulus $G$. In biopolymer networks context, this stiffening plays important roles in many cellular and tissue-level physiological processes, e.g.~it allows the transmission of cellular and tissue forces over large scales, and enables long-range mechanical communication between cells~\cite{GOREN20201152}. Our next goal is to understand the properties of $G$ at $\gamma_{\rm c}$.

In view of the finite jump discontinuity experienced by the shear modulus at $\gamma_{\rm c}$~\cite{robbie_thesis}, we expect $G(\gamma_{\rm c},\kappa)$ to approach a finite value in the $\kappa\!\to\!0$ limit (see Appendix~\ref{sec:appendix_shear_modulus} for additional discussion), i.e.~that
\begin{equation}
\label{eq:G}
 G(\gamma_{\rm c},\kappa\!\to\!0^+) \sim \kappa^{0} \ , 
\end{equation} 
even though the $\kappa\!\to\!0$ limit turns out to be subtle in this context, as discussed below. Next, we aim at calculating the derivative $dG/d\gamma$, evaluated at $\gamma\=\gamma_{\rm c}$, which is nothing but the first nonlinear (shear) elastic coefficient. This observable is sensitive to the presence of low-frequency vibrational modes, and their coupling to external deformation, as pointed out in~\cite{exist} and explained in Appendix~\ref{sec:appendix_shear_modulus}, where it is shown that
\begin{equation}
\label{eq:G_derivative}
 \frac{dG(\gamma,\kappa)}{d\gamma}\biggr\rvert_{\gamma=\gamma_{\rm c}} \!\!\sim \kappa^{-2/3} \ . 
\end{equation}
Equation~\eqref{eq:G_derivative} predicts that $G(\gamma)$ varies strongly with $\gamma$ near $\gamma_{\rm c}$ and that this variation becomes singular as $\kappa\!\to\!0$.

The disordered nature of the elastic networks under discussion, and the corresponding properties of the resulting SSS, imply that such networks deform in a highly non-affine manner as the critical strain-stiffened state as $\gamma_{\rm c}$ is approached. This non-affine response in fact controls the scaling of the first nonlinear elastic coefficient in Eq.~\eqref{eq:G_derivative}, as explained in Appendix~\ref{sec:appendix_shear_modulus}. Similar physical considerations involving low-frequency (soft) vibrational modes and their coupling to the deformation give rise to a scaling relation for the non-affine displacements squared, denoted by $u_{\mbox{\tiny n.a.}}^2$. The latter is shown (see Appendix~\ref{sec:appendix_nonaffine_displacements}) to be related to $\kappa$ as 
\begin{equation}
\label{eq:na_displacements_squared_main}
    u_{\mbox{\tiny n.a.}}^2\!(\gamma_{\rm c},\kappa) \sim \kappa^{-2/3}\,.
\end{equation}

\subsection{The singular perturbation nature\\ of the critical state}

As mentioned above, the perturbation introduced by the soft sub-network to the critical strain-stiffened state is of a singular nature. This fundamental point is further highlighted by comparing the response forces $f_{\mbox{\tiny stiff}}$ to the imposed force perturbation $f_{\mbox{\tiny soft}}$, i.e.~the ratio
\begin{equation}
\label{eq:singular_f}
    f_{\mbox{\tiny stiff}}(\kappa)/f_{\mbox{\tiny soft}}(\kappa) \sim  \kappa ^{-1/3} \ ,
\end{equation}
which diverges in the $\kappa\!\to\!0$ limit. That is, the response is infinitely stronger than the perturbation. 

In addition, let us consider again the shear modulus $G$ at $\gamma_{\rm c}$. A careful calculation, presented in Appendix~\ref{sec:appendix_shear_modulus}, reveals that the value approached in the $\kappa\!\to\!0^+$ limit, differs from $G(\gamma_{\rm c},\kappa=0)$, i.e.~the value obtained at $\kappa\=0$. It is shown that in fact we have
\begin{equation}
\label{eq:singular_G}
    G(\gamma_{\rm c},\kappa\!\to\!0^+) = G(\gamma_{\rm c},\kappa=0) - \Delta{G} \equiv G(\gamma_{\rm c})
\end{equation}
in the limit $\kappa\!\to\!0^+$, where $\Delta{G}\!>\!0$ contains a product of two contributions, one that scales as $\kappa^{-2/3}$ and another as $\kappa^{2/3}$ such that the limit $\kappa\!\to\!0^+$ is finite. Consequently, Eq.~\eqref{eq:singular_G} indicates that the limit $\kappa\!\to\!0^+$ is different from setting $\kappa\=0$, yet again revealing the singular nature of the $\kappa$ perturbation in the problem.

\subsection{Scaling theory beyond the critical point} 

Our next goal is to go beyond the critical point, i.e.~to understand the scaling structure of $G(\gamma,\kappa)$ near $\gamma_{\rm c}$ using Eqs.~\eqref{eq:G}-\eqref{eq:G_derivative}. The first step in achieving this goal is to formulate what is the physical meaning of ``being near $\gamma_{\rm c}$". The latter entails the existence of a characteristic $\kappa$-dependent strain scale $\delta\gamma_*(\kappa)\!>\!0$ such that $|\gamma-\gamma_{\rm c}|\!\ll\!\delta\gamma_*(\kappa)$ defines ``being near $\gamma_{\rm c}$'', both below and above the critical strain $\gamma_{\rm c}$. Consequently, we pose the simplest scaling ansatz of the form
\begin{equation}
\label{eq:G_scaling}
    G(\gamma,\kappa) \sim {\cal F}\!\left(\frac{\gamma_{\rm c}-\gamma}{\delta\gamma_*(\kappa)} \right) \ ,
\end{equation}
where ${\cal F}(\cdot)$ is a dimensionless scaling function. The challenge then is to derive both $\delta\gamma_*(\kappa)$ and ${\cal F}(\cdot)$. 

Denoting $x\!\equiv\!(\gamma_{\rm c}-\gamma)/\delta\gamma_*(\kappa)$ and taking the derivative of $G$ in Eq.~\eqref{eq:G_scaling} with respect to $\gamma$, evaluated at $\gamma_{\rm c}$, we obtain $dG/d\gamma\rvert_{\gamma=\gamma_{\rm c}}\=d{\cal F}/dx\rvert_{x=0}[\delta\gamma_*(\kappa)]^{-1}$. To comply with Eq.~\eqref{eq:G_derivative}, $d{\cal F}/dx\rvert_{x=0}$ has to be finite, i.e.~${\cal F}$ varies linearly with $x$ to leading order, and
\begin{equation}
\label{eq:typical_strain}
    \delta\gamma_*(\kappa) \sim \kappa^{2/3} \ .
\end{equation}
Moreover, Eq.~\eqref{eq:G} implies ${\cal F}(x\=0)\=G(\gamma_{\rm c})$ (the latter is defined in Eq.~\eqref{eq:singular_G}). Consequently, we end up with
\begin{equation}
\label{eq:small_strain_scaling}
   G(\gamma,\kappa)-G(\gamma_{\rm c}) \sim \kappa^{-2/3} (\gamma-\gamma_{\rm c}) \quad \hbox{for} \quad |\gamma-\gamma_{\rm c}|\!\ll\!\kappa^{2/3} \ .
\end{equation}
Equation~\eqref{eq:typical_strain} predicts that the characteristic strain scale $\delta\gamma_*(\kappa)$ vanishes in the limit $\kappa\!\to\!0$ with a nontrivial exponent. Equation~\eqref{eq:small_strain_scaling} predicts that $G(\gamma,\kappa)$ varies linearly with $\gamma$ on top of the constant $G(\gamma_{\rm c})$, both below and above $\gamma_{\rm c}$. That is, we predict that $G(\gamma,\kappa)$ is regular near $\gamma_{\rm c}$ in terms of its variation with the strain $\gamma$. 

Can the scaling form $G(\gamma,\kappa)\!\sim\!{\cal F}\!\left(\frac{\gamma_{\rm c}-\gamma}{\kappa^{2/3}} \right)$ be used to obtain predictions also at smaller strains $\gamma$ below the characteristic strain scale, i.e.~for $\gamma_{\rm c}-\gamma\!\gtrsim\!\kappa^{2/3}$? Below the strain-stiffening transition, the characteristic scale of $G$ is determined by the soft sub-network, i.e.~$G$ has to be linear in $\kappa$. The above scaling form then immediately predicts the $\gamma$ dependence of $G(\gamma,\kappa)$, leading to
\begin{equation}
\label{eq:gamma_3halves}
    G(\gamma,\kappa) \sim \kappa\,(\gamma_{\rm c}-\gamma)^{-3/2} \ ,
\end{equation}
i.e.~we expect ${\cal F}(x)\!\sim\!x^{-3/2}$ in this regime below the critical point, in perfect agreement with the arguments and numerical result of Ref.~\cite{robbie_pre_2018}. Equation~\eqref{eq:gamma_3halves} is expected to hold at an intermediate crossover regime below $\gamma_{\rm c}$, where $G$ significantly increases with $\gamma$. As will be shown below, this prediction is consistent with extensive numerical data available in the literature.

At the same time, it is clear that the scaling ansatz $G(\gamma,\kappa)\!\sim\! {\cal F}\!\left(\frac{\gamma_{\rm c}-\gamma}{\kappa^{2/3}} \right)$ cannot hold in other regimes further away from the critical point, as explained in Appendix~\ref{sec:appendix_widom}. The scaling theory of Eqs.~\eqref{eq:G_scaling}-\eqref{eq:gamma_3halves} also implies that previous attempts~\cite{robbie_nature_physics_2016,mackintosh_prl_2019,fred_arXiv_2022,fred3D_pre_2021} to describe the strain-stiffening transition using a Widom-like scaling~\cite{widom_1965} of the form $G(\gamma,\kappa)\!\sim\!|\gamma-\gamma_{\rm c}|^f{\cal G}_\pm(\kappa/|\gamma-\gamma_{\rm c}|^\phi)$, with $f,\phi\!>\!0$ (here $\pm$ corresponds to the strain regimes above and below $\gamma_{\rm c}$, respectively), cannot be strictly valid, as is further discussed in Appendix~\ref{sec:appendix_widom}.

A similar scaling analysis can be performed for the non-affine displacements squared $u_{\mbox{\tiny n.a.}}^2$, \emph{far} from the critical point, $\gamma_{\rm c}\!-\!\gamma\!\gg\!\delta\gamma_\star(\kappa)$. The analysis is presented in Appendix~\ref{sec:appendix_nonaffine_displacements} and the result reads
\begin{equation}
\label{eq:nonaffine_displacements_away_from_gamma_c}
u_{\mbox{\tiny n.a.}}^2\!(\gamma,\kappa)\sim (\gamma_{\rm c}-\gamma)^{-1}\,.
\end{equation}
Note also that Eq.~\eqref{eq:nonaffine_displacements_away_from_gamma_c} predicts that $u_{\mbox{\tiny n.a.}}^2\!(\gamma,\kappa)$ is independent of $\kappa$ for $\gamma_{\rm c}\!-\gamma\!\gg\!\delta\gamma_\star(\kappa)$. These predictions perfectly agree with the arguments of Ref.~\cite{robbie_pre_2018} and with the numerical simulations presented in~\cite{robbie_thesis}, but stand at odds with the claims of~\cite{mackintosh_prl_2019,fred3D_pre_2021,fred_arXiv_2022}.

\subsection{Comparison to available data obtained in simulations of biopolymer networks models}
\label{sec:agreement}

As mentioned above, the strain-stiffening transition has been quite extensively studied in the literature using numerical simulations of various biopolymer network models in both 2D and 3D. Consequently, it would be interesting to test some of our predictions against numerical results available in the literature. We arrange the discussion according to the order in which the relevant predictions appear in the text.  

{\bf (a) The non-affine displacements} --- We first consider the scaling relation in Eq.~\eqref{eq:na_displacements_squared_main} for the non-affine displacements squared at the critical strain, $u_{\mbox{\tiny n.a.}}^2\!(\gamma_{\rm c},\kappa)$. Recently~\cite{fred_arXiv_2022}, the rheology of fluid-immersed, strain-stiffened networks near $\gamma_{\rm c}$ has been numerically studied in both 2D and 3D using overdamped simulations. A central quantity extracted in these simulation was the so-called ``excess viscosity'', which has been previously shown to be proportional to the square of the non-affine displacements~\cite{asm_pnas_2012}, and hence is directly relevant to the present discussion. In particular, it has been shown in~\cite{fred_arXiv_2022} (cf.~Eq.~5 and Fig.~3c therein) that the excess viscosity at $\gamma_{\rm c}$ scales as $\kappa^{-\xi}$, with $\xi\!\approx\!1.5/2.2\!\approx\!0.68$. This numerical finding is in great quantitative agreement with our prediction in Eq.~\eqref{eq:na_displacements_squared_main}, where we have $\xi\=2/3$. 

It is important to note that the observed scaling in Fig.~3c in~\cite{fred_arXiv_2022} spans $\kappa$ values between $10^{-5}$ and $10^{-3}$. Hence, the agreement with the theoretical prediction --- obtained for $\kappa\!\ll\!1$ without specifying the precise range of validity --- indicates that for the employed networks (with the adopted degree of connectivity $z$~\cite{fred_arXiv_2022}) critical scaling extends to these values of $\kappa$. Moreover, in~\cite{mackintosh_prl_2019} it was found that $\xi\!\approx\!0.677$ for 2D triangular networks (cf.~Fig.~S5b therein) and $\xi\!\approx\!0.668$ for 2D packing-derived networks (cf.~Fig.~S5d therein) over the same range of $\kappa$ values, yet again in excellent quantitative agreement with the prediction in Eq.~\eqref{eq:na_displacements_squared_main}.

{\bf (b) The characteristic strain scale} --- Next, we consider the scaling relation for the characteristic strain scale $\delta\gamma_*(\kappa)$ in Eq.~\eqref{eq:typical_strain}. The latter quantity, as follows from Eqs.~\eqref{eq:G_scaling}-\eqref{eq:gamma_3halves}, characterizes the strain range from both sides of $\gamma_{\rm c}$ where the predicted scaling relations are valid. $\delta\gamma_*(\kappa)$ in Eq.~\eqref{eq:typical_strain} is predicted to shrink with decreasing $\kappa$. Evidence for the existence of such a shrinking strain scale below $\gamma_{\rm c}$ is provided in Fig.~1 of~\cite{robbie_pre_2018}, where $G(\gamma,\kappa)$ is plotted as a function of $\gamma$ for various values of $\kappa$ (note that the parameter varied in that figure is $\kappa^{-1}$), and above $\gamma_{\rm c}$ in Fig.~5.3 of~\cite{robbie_thesis}, where the non-affine displacements squared $u_{\mbox{\tiny n.a.}}^2$ is plotted as a function of $\gamma$ for various values of $\kappa$. While these figures --- reproduced for the sake of completeness in Appendix~\ref{sec:appendix_previous_evidence} --- do not allow to extract the functional form of $\delta\gamma_*(\kappa)$ (to be compared with Eq.~\eqref{eq:typical_strain}), they clearly indicate its existence and the expected trend.   

{\bf (c) The shear modulus} --- Finally, we consider the prediction for the shear modulus $G(\gamma,\kappa)$ in Eq.~\eqref{eq:gamma_3halves}. We find that this prediction is consistent with extensive numerical data available in the literature, as detailed next. As mentioned above, in~\cite{robbie_pre_2018} numerical simulations of strain-stiffening networks in the $\kappa\!\to\!0^+$ limit demonstrate clearly that $G\!\sim\!(\gamma_c\!-\!\gamma)^{-1.5}$ in this limit. Additionally, in~\cite{Rens_JPCB_2016,robbie_nature_physics_2016,mackintosh_prl_2019,fred3D_pre_2021}, it has been observed that $G(\gamma_{\rm c}-\gamma)^{-f}\!\sim\!\kappa(\gamma_{\rm c}-\gamma)^{-\phi}$ (i.e.~$G\!\sim\!\kappa(\gamma_{\rm c}-\gamma)^{f-\phi}$) with $f\!-\!\phi$ close to $-3/2$, below $\gamma_{\rm c}$. Specifically, in~\cite{Rens_JPCB_2016}, $f\!-\!\phi\!\approx\!0.53\!-\!2.0\=-1.47$ (2D undistorted honeycomb lattice, cf.~Fig.~4a therein) and $f\!-\!\phi\!\approx\!0.63\!-\!1.9\=-1.27$ (2D undistorted triangular lattice, cf.~Fig.~4d therein) have been reported; in~\cite{robbie_nature_physics_2016}, $f\!-\!\phi\!\approx\!0.75\!-\!2.1\=-1.35$ (2D triangular lattice, cf.~Fig.~2b therein), $f\!-\!\phi\!\approx\!0.84\!-\!2.2\=-1.36$ (2D Mikado network, cf.~Fig.~2b therein) and $f\!-\!\phi\!\approx\!0.8\!-\!2.2\=-1.4$ (3D fcc lattice, cf.~Fig.~2b therein) have been reported; in~\cite{mackintosh_prl_2019}, $f\!-\!\phi\!\approx\!0.73\!-\!2.26\=-1.53$ (2D triangular network, cf.~Fig.~3a therein) and $f\!-\!\phi\!\approx\!0.68\!-\!2.05\=-1.37$ (2D packing-derived network, cf.~Fig.~3b therein) have been reported; and in~\cite{fred3D_pre_2021}, $f\!-\!\phi\!\approx\!0.86\!-\!2.6\!=\!-1.74$ (3D packing-derived networks with $z\!=\!4.0$, cf.~Fig.~S11 therein), $f\!-\!\phi\!\approx\!0.79\!-\!2.5\!=\!-1.71$ (3D packing-derived networks with $z\!=\!3.3$, cf.~Figs.~2-3 therein), and $f\!-\!\phi\!\approx\!0.92\!-\!2.8\!=\!-1.88$ (3D random geometric graphs, cf.~Fig.~S8 therein) have been reported. These are all in reasonable agreement with our prediction $f\!-\!\phi\=-3/2\=-1.5$ in Eq.~\eqref{eq:gamma_3halves}. 

The observations discussed above provide independent support to our scaling theory. The diversity of the fibrous network models employed, in both 2D and 3D, provides support to the generality (and dimension independence) of the developed framework in which the entire network is viewed as two interacting sub-networks, where the soft (here corresponding to fiber bending) sub-network is treated as a perturbation to the stiff (here corresponding to fiber stretching/compression) sub-network near $\gamma_{\rm c}$. Moreover, the comparison to the literature data provides some estimates for the range of $\kappa$ values for which critical scaling might be observed. Since our current theory does not predict this range, and its dependence on $z$ and other potentially relevant physical factors, future work should further address this issue.  

We also note that in~\cite{merkel_pnas_2019} (cf.~Fig.~4D therein) and in~\cite{robbie_thesis} (cf.~Fig.~5.1 therein), $G-G(\gamma_{\rm c})\!\sim\!(\gamma-\gamma_{\rm c})$ has been observed above and close to $\gamma_{\rm c}$ for $\kappa\=0$, apparently in agreement with the prediction in Eq.~\eqref{eq:small_strain_scaling}. This apparent agreement, however, should be taken with caution as Eq.~\eqref{eq:small_strain_scaling} is valid in the $\kappa\!\to\!0$ limit (and note the predicted diverging $\kappa^{-2/3}$ prefactor), which may (or may not) differ from its $\kappa\=0$ counterpart in terms of the scaling with $\gamma$. Future work should further clarify this point as well.

\section{Discussion and outlook}

In this work, we developed a comprehensive scaling theory of the strain-stiffening transition in disordered elastic networks, at and near the critical strain $\gamma_{\rm c}$. Building on the rigidity scales separation typically featured by such networks, we treated the sub-network of weak forces as a random and isotropic perturbation of dimensionless magnitude $\kappa\!\ll\!1$ applied to the anisotropic stiff sub-network at $\gamma_{\rm c}$. With this conceptual framework in mind, we theoretically predicted the dependence of the salient physical quantities in the problem on $\kappa$ at the critical strain $\gamma_{\rm c}$. We expect our scaling theory to apply to systems in any spatial dimension $\dbar\!\ge\!2$, as also supported by its quantitative agreement with existing numerical results in both 2D and 3D, as spelled out in Sect.~\ref{sec:agreement}.

The existence of floppy modes that are uncoupled (orthogonal) to the applied strain have been shown to dominate the stiff sub-network's response to the weak forces, which is characterized by quartic anharmonicity, in line with recent theoretical arguments~\cite{manning_rigidity1_pre_2022,manning_rigidity2_pre_2022}, and similarly to observations in gently compressed packings of nonspherical particles~\cite{elipse_packings_prl_2009,ohern_pre_quartic_modes_2018,ohern_deformable_particles_2021}. The $\kappa$ scaling of the network's node displacements and of its energy at $\gamma_{\rm c}$ then follow, cf.~Eqs.~\eqref{eq:displacement}-\eqref{eq:energy}. Furthermore, analyzing the state-of-self-stress (SSS) and its breakdown in the presence of weak interactions, as well as the accompanying non-affinity, allowed us to predict the $\kappa$ scaling of the forces in the problem and of the macroscopic modulus $G$ at $\gamma_{\rm c}$, cf.~Eqs.~\eqref{eq:SSS}-\eqref{eq:na_displacements_squared_main}. The structure of the theory highlights the role of $\kappa$ as a singular perturbation applied to the critical strain-stiffened state, two manifestations of which are discussed in the context of Eqs.~\eqref{eq:singular_f}-\eqref{eq:singular_G}

With the complete $\kappa$ scaling theory at $\gamma_{\rm c}$, we then extended the theory beyond the critical state and derived scaling relations for the macroscopic modulus $G(\gamma,\kappa)$ in Eqs.~\eqref{eq:G_scaling}-\eqref{eq:gamma_3halves}, and for the squared nonaffine displacements $u_{\mbox{\tiny n.a.}}^2(\gamma,\kappa)$ in Eq.~(\ref{eq:nonaffine_displacements_away_from_gamma_c}). These scaling relations highlight the existence of a previously unidentified characteristic $\kappa$-dependent strain scale $\delta\gamma_*(\kappa)$ near $\gamma_{\rm c}$. Available numerical results in the literature lend independent support to some of the predictions, which should be further tested in their entirety in future work. Furthermore, the structure of the emerging scaling theory indicates that the Widom-like scaling form assumed in previous work~\cite{robbie_nature_physics_2016, mackintosh_prl_2019} cannot be strictly and self-consistently valid. 

The progress made in this work also opens the way for additional future investigations. Immediate and pressing ones are the substantiation of the predicted strain scale in Eq.~\eqref{eq:typical_strain} and of the linear $\gamma$ scaling near $\gamma_{\rm c}$ according to Eq.~\eqref{eq:small_strain_scaling}. In addition, the validity of the conceptual procedure in which a critical $\gamma_{\rm c}$ state is supplemented with a $\kappa\!\ll\!1$ soft sub-network should be further tested in comparison to a continuous straining procedure at $\kappa\!>\!0$. Yet another important direction for future investigation is understanding the range of the critical scaling, which is currently not predicted by our theory. One is particularly interested in the effect of the stiff sub-network's coordination $z$ on it, and on other observables discussed in this work. Very preliminary, high accuracy, single realization computer simulations indeed indicate such $z$ dependence~\cite{lerner2022scaling}.  

In this work, we did not address spatial aspects of the strain-stiffening transition. Yet, it is conceivable that a characteristic lengthscale $\ell_\kappa$ that increases with decreasing $\kappa$ exists in this problem~\cite{gustavo_pre_2014,robbie_pre_2018}; the results from~\cite{gustavo_pre_2014,robbie_pre_2018} lead to the expectation that $\ell_\kappa\!\sim\!\kappa^{-1/3}$ should be observed~\cite{footnote2}.
In this context, it would be interesting to explore finite-size effects encountered in computer simulations and their dependence on spatial dimension.

As discussed above and in Appendix~\ref{sec:appendix_widom}, the scaling form in Eq.~\eqref{eq:G_scaling} cannot remain valid away from the critical point $\gamma_{\rm c}$, i.e.~different scaling relations are expected for $|\gamma-\gamma_{\rm c}|\!\gg\!\delta\gamma_*(\kappa)$. While these scaling relations have been numerically explored~\cite{robbie_thesis,mackintosh_pre_2016,mackintosh_prl_2019,robbie_nature_physics_2016,fred_arXiv_2022}, theoretically deriving them remains a challenge for future work. Finally, we did not consider in this work possible differences between the effect of applied shear and dilatational strains, and in particular all numerical validation tests were performed under shear strains. It would be interesting to systematically explore dilatational strains in future work. 

\acknowledgements

We thank Gustavo D\"uring for enlightening discussions that led to this work and Eric Lerner for his assistance with the graphics of Fig.~1. E.L.~acknowledges Support from the NWO (Vidi grant no.~680-47-554/3259). E.B.~acknowledges support from the Ben May Center for Chemical Theory and Computation and the Harold Perlman Family.

\appendix




\section{The geometric operator and states-of-self-stress}
\label{sec:appendix_sss_operator}

Consider a disordered network of edges and nodes, similar to that shown in Fig.~1a of the main text. States-of-self-stress (SSS) are assignment of putative (scalar) edge-forces $\{f_{ij}\}$ such that, for each node $i$
\begin{equation}\label{eq:SSS_definition}
    \sum_{\mbox{\tiny neighbors } j(i)}\nv_{ji}f_{ij} = \zerovector\,,
\end{equation}
where $\nv_{ji}$ is the unit vector pointing from node $j$ to node $i$, the sum runs over neighbors $j$ connected \emph{by an edge} to node $i$, and we choose the convention that positive (negative) $f_{ij}$'s represent compressive (tensile) forces. Generally, an isotropic disordered network must have $z\!\ge\!2\dbar$ for at least one SSS to exist; however, floppy networks with $z\!<\!2\dbar$ can develop a SSS, as indeed happens in the strain-stiffening problem.

At this point, it is convenient to adopt a bra-ket notation; in this notation, Eq.~(\ref{eq:SSS_definition}) takes the form
\begin{equation}
    {\cal S}^T\ket{f} = \ket{0}\,,
\end{equation}
where the operator ${\cal S}$ represents the geometry of the network. It is defined as
\begin{equation}\label{eq:es_definition}
    \calBold{S}_{ij,k} = \frac{\partial r_{ij}}{\partial \xv_k} = (\delta_{jk}-\delta_{ik})\nv_{ij}\,,
\end{equation}
where $r_{ij}$ is the pairwise distance between nodes $i,j$, $\delta_{jk}$ is the Kronecker delta, $\xv_k$ is the position vector of the $k$'th node, and $\nv_{ij}\!\equiv\!(\xv_j\!-\!\xv_i)/r_{ij}$ is the unit vector pointing from node $i$ to node $j$. For convenience, we define the concatenated operator ${\cal S}{\cal S}^T$ such that SSS correspond to its eigenmodes associated with zero eigenvalues, denoted here and in what follows by $\lambda$.

\subsection{The variation $\delta\lambda(\kappa)$}
\label{sec:appendix_sss_variation}

In the absence of soft interactions (corresponding to $\kappa\!=\!0)$, strain-stiffened networks feature a single SSS at the critical strain $\gamma_{\rm c}$. We denote this SSS by $\ket{\phi}$ (note that above, a general SSS was denoted by $\ket{f}$), and recall that ${\cal S}^T\ket{\phi}\!=\!\ket{0}$; how does this SSS break down under the introduction of soft interactions into the network? Let us assume that by introducing bending energies of order $\kappa$, the nodes undergo displacements $\uv_\star$ (of typical magnitude $u_\star(\kappa)$, see Fig.~1 in the main text). How does $\lambda\!=\!\bra{\phi}{\cal S}{\cal S}^T\ket{\phi}$ --- where ${\cal S}$ is defined only on the stiff subsystem --- change under node-displacements $\uv_\star$? Denoting by $\delta S$, $\delta\lambda$ and $\ket{\delta\phi}$ the variations of ${\cal S}$, $\lambda$ and $\ket{\phi}$ under the displacements $\uv_\star$, respectively, we obtain to first order 
$\delta\lambda\=2\bra{\phi}{\cal S}{\cal S}^T\ket{\delta\phi} + 2\bra{\phi} \delta {\cal S}{\cal S}^T\ket{\phi}\=0$, because ${\cal S}^T\ket{\phi}\!=\!\ket{0}$. The next order, which involves contributions of the form $\bra{\phi}\delta{\cal S}\delta{\cal S}^T\ket{\phi}$ does not vanish, hence we conclude that to leading order we have
\begin{equation}
\delta\lambda \sim u_\star^2 \sim \kappa^{2/3}\,.
\end{equation}

\section{Linear and first nonlinear shear moduli}
\label{sec:appendix_shear_modulus}

For any a solid (i.e.~a system having a finite shear modulus) at zero temperature, given a potential energy $U$, the athermal linear shear modulus reads~\cite{lutsko}
\begin{equation}\label{eq:G_exact_expression}
    G = G_{\mbox{\tiny Born}} - \Fv_\gamma\cdot\calBold{H}^{-1}\cdot\Fv_\gamma\,,
\end{equation}
where $G_{\mbox{\tiny Born}}\!\equiv\!V^{-1}\partial^2U/\partial\gamma^2$ is the affine contribution to the shear modulus (which is regular), $\Fv_\gamma\!\cdot\!\calBold{H}^{-1}\!\cdot\!\Fv_\gamma$ is the relaxation term that accounts for the softening due to non-affine displacements, $\Fv_\gamma\!\equiv\!-\frac{\partial^2U}{\partial\gamma\partial\xv}$ are the forces that emerge from applying an affine deformation, $\calBold{H}\!\equiv\!\frac{\partial^2U}{\partial\xv\partial\xv}$ is the Hessian matrix of the potential energy, and $\gamma$ parametrizes the strain tensor $\calBold{\epsilon}$ as
\begin{equation}
    \calBold{\epsilon} = 
    \frac{1}{2}\left( 
    \begin{array}{cc}
    0 & \gamma \\ \gamma & \gamma^2
    \end{array}\right)\,.
\end{equation}

It has been shown~\cite{matthieu_thesis} that the shear modulus of a relaxed Hookean spring network at $U\!=\!0$, in which all spring stiffnesses are set to unity, is given by
\begin{equation}\label{eq:G_mw}
G = \frac{1}{V} \sum_{\mbox{\tiny SSS }\ell} \braket{\phi_\ell}{\sFrac{\partial r}{\partial\gamma}}^2\,,
\end{equation}
where $\ket{\phi_\ell}$ is the $\ell^{\mbox{\scriptsize th}}$ SSS, and $\ket{\frac{\partial r}{\partial\gamma}}$ is a vector in which each component corresponds to the derivative $\partial r_{ij}/\partial\gamma$, with $r_{ij}$ denoting the pairwise distance pertaining to a particular edge $i,j$ of the network. Isotropic, underconstrained (sub-isostatic) networks feature no SSS, and thus $G\!=\!0$ as observed. Anisotropic strain-stiffened networks, however, feature a single SSS $\ket{\phi}$ at the critical strain $\gamma_{\rm c}$. The latter state features a finite coupling to the strain that scales with the system size as $\braket{\phi}{\frac{\partial r}{\partial\gamma}}^2\!\sim\!N$ (with a $z$-dependent prefactor~\cite{robbie_pre_2018}, and notice that we adopt the convention $\braket{\phi}{\phi}\!=\!1$). Consequently, strain-stiffened networks feature an intensive shear modulus of order $G\!\sim\!{\cal O}(1)$ in the absence of bending energy ($\kappa\!=\!0$, and recall that the stiff bonds rigidity has been set to unity). 

How does the shear modulus $G$ of critical states ($\gamma\!=\!\gamma_{\rm c}$) change upon introducing the weak interactions $U_{\mbox{\tiny weak}}$ featuring stiffnesses and forces of order $\kappa$? We consider Eq.~(\ref{eq:G_exact_expression}) and first point out that while $G_{\mbox{\tiny Born}}$ is regular~\cite{lutsko}, the second term in Eq.~(\ref{eq:G_exact_expression}) --- containing a full contraction of the \emph{inverse} of the Hessian $\calBold{H}^{-1}$ --- can be singular by virtue of low-lying vibrational modes of $\calBold{H}$. We review the vibrational spectrum of strain-stiffened networks next.

\subsection{Vibrational spectrum at the critical state}
\label{sec:vibrational_spectrum}

Consider first the $\kappa\!=\!0$ case, namely in the absence of soft interactions. Strain-stiffened networks at $\gamma\!=\!\gamma_{\rm c}$ are expected to have $\sim\! N(2\dbar\!-\!z)$ zero modes, since we consider underconstrained networks with $z\!<\!2\dbar$. Zero modes are putative displacements $\Psiv$ that satisfy
\begin{equation}
    \nv_{ij}\cdot(\Psiv_j - \Psiv_i) = 0\,,
\end{equation}
for all (stiff) bonds connecting pairs of nodes $i,j$. The above equation tell us that zero modes in $\kappa\!=\!0$ networks do not deform any of the (stiff) bonds. At the strain-stiffening transition the energy is vanishingly small, therefore the Hessian takes the form
\begin{equation}
    \calBold{H}_{k\ell}(\gamma_{\rm c},\kappa\!=\!0) = \!\!\sum_{\mbox{\tiny bonds } i,j}\!\!\!\!\Lambda_{ij,k\ell}\,\,\, \nv_{ij}\otimes\nv_{ij}\,,
\end{equation}
where $\Lambda_{ij,k\ell}\!\equiv\!(\delta_{jk}\!-\!\delta_{ik})(\delta_{j\ell}\!-\!\delta_{i\ell})$, $\otimes$ represents an outer (dyadic) product, and we take the stiffness of the stiff bonds to be unity. 

Once the soft interactions are introduced to the $\kappa\!=\!0$ strain-stiffened network, and after the minimization of the total energy $U\!=\!U_{\mbox{\tiny stiff}}+U_{\mbox{\tiny weak}}$ during which nodes are displaced a typical distance $u_\star$, forces of order $f_{\mbox{\tiny stiff}}\!\sim\!\kappa^{2/3}$ emerge in the stiff sub-network, as explained in the main text (see discussion preceding Eq.~(4) in the main text). The Hessian now assumes the form
\begin{eqnarray}\label{eq:finite_kappa_hessian}
    \calBold{H}_{k\ell}(\gamma_{\rm c},\kappa) & = & \bigg[\sum_{\mbox{\tiny stiff bonds } i,j}\!\!\!\!\!\!\!\Lambda_{ij,k\ell}\, \nv_{ij}\otimes\nv_{ij} \nonumber \\
    &&-\!\!\!\!\!\!\!\!\sum_{\mbox{\tiny stiff bonds } i,j}\!\!\!\!\!\!\!\!\Lambda_{ij,k\ell}\, \frac{f_{ij}}{r_{ij}}\big( \calBold{I}-\nv_{ij}\otimes\nv_{ij}\big)\bigg] + \calBold{H}_{\mbox{\tiny soft}} \nonumber \\
    & \equiv & \calBold{H}_1 + \calBold{H}_2 + \calBold{H}_{\mbox{\tiny soft}}\,, 
\end{eqnarray}
where $\calBold{H}_{\mbox{\tiny soft}}\!\equiv\!\frac{\partial^2U_{\mbox{\tiny soft}}}{\partial\xv\partial\xv}\!\sim\!\kappa$. What frequency do the zero modes $\Psiv$ in the $\kappa\!=\!0$ (critical) network at $\gamma_{\rm c}$ acquire in the perturbed, $\kappa\!>\!0$ network? Employing standard degenerate perturbation theory (see e.g.~\cite{phonon_widths}), the new frequencies squared are eigenvalues of the matrix
\begin{equation}
    \widetilde{\calBold{H}}_{\ell m} \equiv \Psiv^{(\ell)}\cdot\delta\calBold{H}\cdot\Psiv^{(m)} = \Psiv^{(\ell)}\cdot(\delta\calBold{H}_1 + \delta\calBold{H}_2) \cdot\Psiv^{(m)} + {\cal O}(\kappa) \,,
\end{equation}
where $\delta\calBold{H}\!=\!\calBold{H}(\kappa)\!-\!\calBold{H}(\kappa\!=\!0)$, and $\Psiv^{(\ell)},\Psiv^{(m)}$ are degenerate zero modes of the unperturbed network. Since 
\begin{equation}
\delta\nv_{ij}=\frac{\partial\nv_{ij}}{\partial\xv}\cdot\uv_\star\,,
\end{equation}
then
\begin{equation}
    \Psiv^{(\ell)}\cdot\delta\calBold{H}_1\cdot\Psiv^{(m)} \sim u_\star^2 \sim \kappa^{2/3}\,.
\end{equation}
Next, since the forces in the stiff, perturbed sub-network scale as $f_{\mbox{\tiny stiff}}\!\sim\!\kappa^{2/3}$, then also
\begin{equation}
    \Psiv^{(\ell)}\cdot\delta\calBold{H}_2\cdot\Psiv^{(m)} \sim f_{\mbox{\tiny stiff}} \sim \kappa^{2/3}\,.
\end{equation}
We conclude that the characteristic scale $\omega_\kappa$ of the newly acquired frequencies of the soft modes in the perturbed network is dominated by the stiff sub-network, and follows
\begin{equation}\label{eq:freq_scaling}
    \omega_\kappa \sim \kappa^{1/3}\,.
\end{equation}
Assuming that, in the perturbed system, the two terms $\calBold{H}_1$ and $\calBold{H}_2$ contribute equally to the frequencies $\omega_\kappa$ of soft modes $\Psiv$, one then expects that for stiff bonds connecting nodes $i$ and $j$
\begin{equation}\label{eq:radial_projections_finite_kappa}
    \nv_{ij}\cdot(\Psiv_j - \Psiv_i) \sim \omega_\kappa \sim \kappa^{1/3}\,.
\end{equation}
This scaling with frequency of radial projections of soft modes resembles that found in other jamming problems, see e.g.~\cite{silbert_pre_2016}. 

\subsection{Relaxation term of the shear modulus}

With the results in Eqs.~(\ref{eq:freq_scaling}) and (\ref{eq:radial_projections_finite_kappa}) at hand, we are now in position to analyze the relaxation term of the shear modulus (the second term in the RHS of Eq.~(\ref{eq:G_exact_expression})). Using the eigenbasis of the Hessian $\{\Psiv^{(\ell)}\}_{\ell=0}^{N\dbar}$, and their associated eigenvalues $\omega_\ell^2$, we write
\begin{equation}
    \Fv_\gamma\cdot\calBold{H}^{-1}\cdot\Fv_\gamma = \sum_\ell \frac{\big(\Fv_\gamma\cdot\Psiv^{(\ell)}\big)^2}{\omega_\ell^2}\,. 
\end{equation}
This sum could be dominated by the soft modes $\Psiv$ (which were \emph{zero} modes in the unperturbed, $\kappa\!=\!0$ network), with associated frequencies $\omega_\kappa\!\sim\!\kappa^{1/3}$. Notice, however, that $\Fv_\gamma$ is dominated by the stiff potential $U_{\mbox{\tiny stiff}}$, and so we can write
\begin{equation}
    (\Fv_\gamma)_i \simeq \sum_{\mbox{\tiny stiff  neighbors } j(i)} \frac{\partial r_{ij}}{\partial\gamma}\nv_{ji}\,,
\end{equation}
or, in our bra-ket notation (and see Eq.~(\ref{eq:es_definition}))
\begin{equation}\label{eq:foo00}
    \ket{F_\gamma} = {\cal S}^T\ket{\sFrac{\partial r}{\partial\gamma}}\,,
\end{equation}
where, once again, ${\cal S}$ is only defined on the stiff sub-network. Following Eq.~(\ref{eq:foo00}), the overlaps squared $(\Fv_\gamma\cdot\Psiv)^2$ for soft modes $\Psiv$ can be written as
\begin{equation}\label{eq:overlaps_squared}
   \big(\Fv_\gamma\cdot\Psiv\big)^2 = \bra{\sFrac{\partial r}{\partial\gamma}}{\cal S}\ket{\Psi}^2 \sim \omega_\kappa^2 \sim \kappa^{2/3}\,,
\end{equation}
and notice we have used Eq.~(\ref{eq:radial_projections_finite_kappa}), namely that $\bra{\Psi}{\cal S}^T{\cal S}\ket{\Psi}\!\sim\!\kappa^{2/3}$. We conclude from this discussion that the contribution of soft modes $\Psiv$ to the relaxation term of the shear modulus $G$ follows
\begin{equation}
    \frac{\big(\Fv_\gamma\cdot\Psiv\big)^2}{\omega_\kappa^2} \sim \frac{\kappa^{2/3}}{\kappa^{2/3}}\sim {\cal O}(1)\,.
\end{equation}
Finally, these considerations lead us to expect that $G\!\sim\!\kappa^0$, as indeed shown in Fig.~3a of the main text. The above results also explain why the shear modulus jumps discontinuously from $\kappa\!\to\!0^+$ to $\kappa\!=\!0$.

\subsection{The first nonlinear shear modulus}

The first nonlinear shear modulus $dG/d\gamma$ can be shown to consist of 4 terms containing zero, one, two, and three contractions of the \emph{inverse} of the Hessian $\calBold{H}^{-1}$~\cite{athermal_elasticity_pre_2010}. The most singular of these terms can be expressed via the eigenbasis of the Hessian as~\cite{exist}
\begin{widetext}
\begin{equation}\label{eq:dG_dgamma_most_singular_term}
    \frac{dG}{d\gamma} \simeq \frac{1}{V}\sum_{\ell m n}\frac{(\Psiv_\ell\cdot\Fv_\gamma)(\Psiv_m\cdot\Fv_\gamma)(\Psiv_n\cdot\Fv_\gamma)\,(\calBold{U}'''\!\tripleCdot\Psiv_\ell\Psiv_m\Psiv_n)}{\omega_\ell^2\omega_m^2\omega_n^2}+ {\cal O}(\calBold{H}^{-2})\,,
\end{equation}
\end{widetext}
where $\calBold{U}'''\!\equiv\!\frac{\partial^3U}{\partial\xv\partial\xv\partial\xv}$ is the rank-3 tensor of derivatives of the potential with respect to coordinates $\xv$, and $\tripleCdot$ denotes a triple contraction. In order to resolve the $\kappa$-scaling of $dG/d\gamma$, we are left with analyzing the contribution of soft modes $\Psiv$ to $dG/d\gamma$, and in particular the full contractions
\begin{widetext}
\begin{equation}
    \calBold{U}'''\!\tripleCdot\Psiv\Psiv\Psiv \sim\,\calBold{U}'''\big|_{\kappa=0}\tripleCdot\Psiv\Psiv\Psiv + \frac{\partial^4 U}{\partial\xv\partial\xv\partial\xv\partial\xv}\bigg|_{\kappa=0}::\uv_\star\Psiv\Psiv\Psiv \sim u_\star \sim \kappa^{1/3}\,,
\end{equation}
\end{widetext}
where $::$ denotes a quadruple contraction, and notice that the cubic anharmonicity $\calBold{U}'''\big|_{\kappa=0}\!\tripleCdot\Psiv\Psiv\Psiv$ is expected to be vanishingly small from stability considerations, as explained in the main text. Together with the previously established scaling of the overlaps $\Psiv\cdot\Fv_\gamma\!\sim\!\kappa^{1/3}$ (see Eq.~(\ref{eq:overlaps_squared}) above), the scaling of the soft frequencies $\omega_\kappa\!\sim\!\kappa^{1/3}$ (see Eq.~(\ref{eq:freq_scaling}) above), and the form of  Eq.~(\ref{eq:dG_dgamma_most_singular_term}), we conclude that 
\begin{equation}
    \frac{dG}{d\gamma} \sim \kappa^{-2/3}\,.
\end{equation}
\begin{figure*}[ht!]
 \includegraphics[width = 1\textwidth]{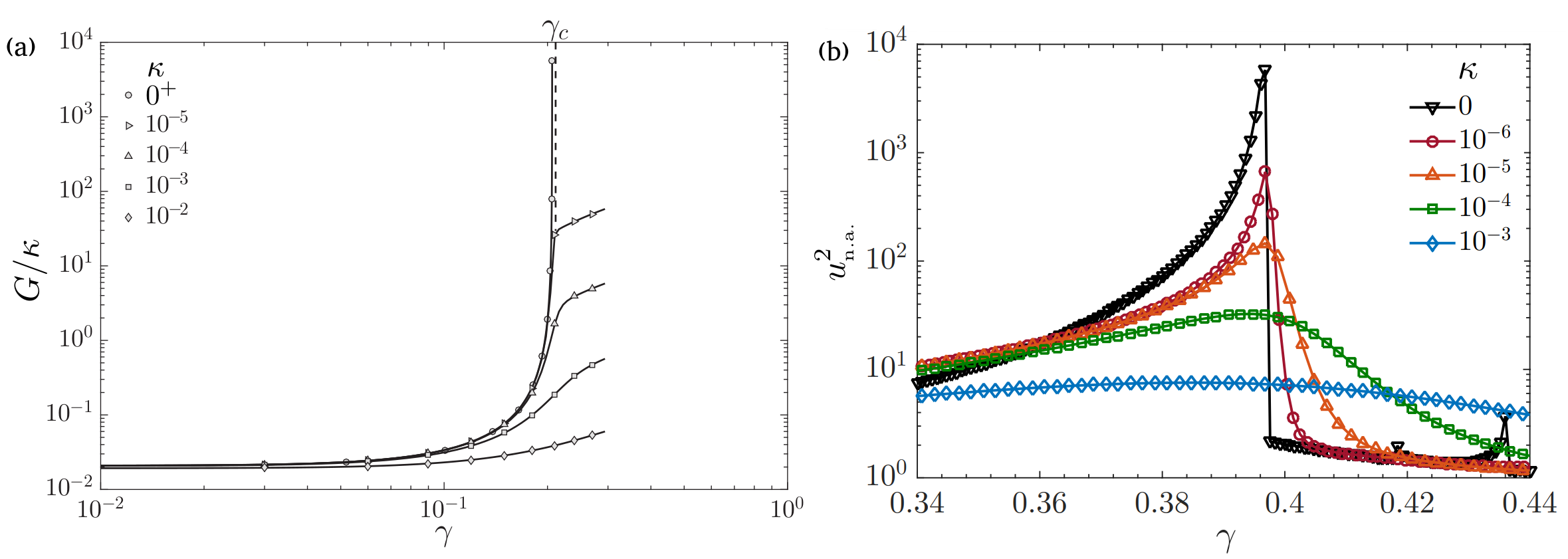}
  \caption{\footnotesize (a) Adapted from~\cite{robbie_pre_2018}. The shear modulus $G$ --- expressed in terms of the soft stiffness $\kappa$ --- is plotted against the shear strain $\gamma$. Here $G$ was measured in packing-derived disordered networks with soft bending energies of stiffness $\kappa$, for various values of $\kappa$. The $\kappa\!\to\!0^+$ data is generated using the geometric formalism introduced in~\cite{robbie_pre_2018}, and see further details therein. The strain scale $\delta\gamma_\star(\kappa)$ can be seen as the strain at which the finite-$\kappa$ signals depart from the $\kappa\!\to\!0^+$ signal. (b) Adapted from~\cite{robbie_thesis}. The non-affine displacements squared (cf.~Eq.~\eqref{eq:na_displacements_squared_main} in the main text, which appears again in Eq.~(\ref{eq:na_displacements_squared})) are plotted against the shear strain $\gamma$, across the strain-stiffening transition. Here measurements were performed in diluted honeycomb lattices with $z\!=\!2.73$, see further details in~\cite{robbie_thesis}.} 
  \label{fig:fig2_SI}
\end{figure*}

\section{Non-affine displacements of strain-stiffened networks}
\label{sec:appendix_nonaffine_displacements}

The non-affine displacements squared can be written using the eigenbasis decomposition of the Hessian as~\cite{athermal_elasticity_pre_2010}
\begin{equation}\label{eq:na_displacements_squared}
    u_{\mbox{\tiny n.a.}}^2 = \sum_\ell \frac{(\Fv_\gamma\cdot\Psiv_\ell)^2}{\omega_\ell^4} \sim \frac{\kappa^{2/3}}{\kappa^{4/3}}\sim \kappa^{-2/3}\,,
\end{equation}
where we have again used Eqs.~(\ref{eq:freq_scaling}) and (\ref{eq:overlaps_squared}) for soft modes $\Psiv$ and their associated eigenfrequencies $\omega_\kappa$. This prediction, which appears in Eq.~\eqref{eq:na_displacements_squared_main} in the main text, is in very good agreement with the numerical simulations of~\cite{fred_arXiv_2022}, as is also discussed in the main text. In the context of~\cite{fred_arXiv_2022}, we note that under overdamped conditions (mimicking fluid-immersed networks), we expect the excess viscosity of the system (i.e.~the viscosity on top of that of the surrounding fluid) to scale with the non-affine displacements squared $u_{\mbox{\tiny n.a.}}^2$, as argued in~\cite{asm_pnas_2012,gustavo_pre_2014}. 

We can next employ the emergent strain scale $\delta\gamma_\star(\kappa)$ (cf.~Eq.~(\ref{eq:typical_strain})) and the prediction for the scaling $u_{\mbox{\tiny n.a.}}^2\!\sim\!\kappa^{-2/3}$ at the critical point (given by Eq.~\ref{eq:na_displacements_squared}), to predict the behavior of $u_{\mbox{\tiny n.a.}}^2$ \emph{away} from the critical point, namely at strains $\gamma_{\rm c}-\gamma\!\gg\!\delta\gamma_\star$. To this aim we propose the scaling ansatz 
\begin{equation}
u_{\mbox{\tiny n.a.}}^2\!(\gamma,\kappa)\sim\kappa^{-2/3}\,{\cal 
Y}\!\left(\frac{\gamma_{\rm c}\!-\!\gamma}{\delta\gamma_*(\kappa)} \right)\,,
\end{equation}
where ${\cal Y}(x)$ with $x\!\equiv\!(\gamma_{\rm c}\!-\!\gamma)/\kappa^{2/3}$ is a scaling function whose form is determined as follows. First, satisfying the scaling $u_{\mbox{\tiny n.a.}}^2\!\sim\!\kappa^{-2/3}$ at the critical point $\gamma_{\rm c}$ (corresponding to $x\!=\!0$) requires ${\cal Y}(0)\!\sim\!\mbox{const}$. On the other hand, for strains far away from the critical point (i.e.~at $x\!\gg\!1$) we expect $u_{\mbox{\tiny n.a.}}^2$ to become $\kappa$-independent (as $\kappa\!\to\!0^+$), as can be gleaned from the small-strain behavior of $u_{\mbox{\tiny n.a.}}^2$ in Fig.~\ref{fig:fig2_SI}b. Losing the $\kappa$-dependence at large $x$ requires that ${\cal Y}(x)\!\sim\!1/x$ for $x\!\gg\!1$. From here, we deduce the scaling form of non-affine displacements away from the critical point as
\begin{equation}
u_{\mbox{\tiny n.a.}}^2\!(\gamma,\kappa) \sim(\gamma_c-\gamma)^{-1}\quad\mbox{for}\quad \gamma_{\rm c}-\gamma \gg \delta\gamma_\star\,,
\end{equation}
in perfect agreement with the scaling predictions of~\cite{robbie_pre_2018} and with the numerical simulations presented in~\cite{robbie_thesis}.

\section{Relation to the Widom-like scaling ansatz}
\label{sec:appendix_widom}

Previous numerical work \cite{fred_arXiv_2022,robbie_nature_physics_2016,mackintosh_prl_2019} attempted to describe the strain-stiffening transition using a Widom-like scaling \cite{widom_1965} of the form
\begin{equation}
    G(\gamma,\kappa) \sim |\gamma-\gamma_{\rm c}|^f {\cal G}_{\pm}\left(
    \frac{\kappa}{|\gamma - \gamma_{\rm c}|^\phi}\right)\,,
\end{equation}
where $\pm$ corresponds to the strain regimes above and below $\gamma_{\rm c}$, respectively. This scaling form is consistent with our
scaling results in Eqs.~(\ref{eq:G_scaling})-(\ref{eq:typical_strain}) in the main text, if one sets $f\!=\!0$ and $\phi\!=\!3/2$, and properly maps ${\cal G}_{\pm}(\cdot)$ to ${\cal F}(\cdot)$ of
Eq.~(\ref{eq:G_scaling}). However, the work of \cite{fred_arXiv_2022,robbie_nature_physics_2016,mackintosh_prl_2019} argued that $f,\phi\!>\!0$, which is evidently inconsistent with our findings. The source of the problem is that the continuity of ${\cal G}_{\pm}(s)$ at $\gamma_{\rm c}$ has been invoked to obtain ${\cal G}_{\pm}\!\sim\!s^{f/\phi}$, which leads to $G(\gamma_{\rm c}, \kappa)\!\sim\!\kappa^{f/\phi}$, clearly contradicting Eq.~(\ref{eq:singular_G}) in the main text. Carefully looking at the analysis of \cite{fred_arXiv_2022,robbie_nature_physics_2016,mackintosh_prl_2019}, it is evident that the prediction $G(\gamma_{\rm c}, \kappa)\!\sim\!\kappa^{f/\phi}$ has been compared to data at $\gamma_{\rm c}$ at relatively large $\kappa$ values (see, for example, the insets of Fig.~3 of \cite{mackintosh_prl_2019}), apparently above the critical scaling
regime described by Eq.~(\ref{eq:singular_G}) in the main text. 

\section{Previous numerical evidence for the existence of {\large $\delta\gamma_\star(\kappa)$}}
\label{sec:appendix_previous_evidence}

In Fig.~\ref{fig:fig2_SI} above, we reproduce Fig.~1 from~\cite{robbie_pre_2018} (panel (a)) and Fig.~5.3 from~\cite{robbie_thesis} (panel (b)), indicating the existence of a strain scale $\delta\gamma_\star(\kappa)$ that vanishes as $\kappa\!\to\!0^+$, both at strains below the critical strain, $\gamma\!<\!\gamma_{\rm c}$ (panel (a)), and at strains above the critical strain, $\gamma\!>\!\gamma_{\rm c}$ (panel (b)).


\begin{thebibliography}{40}
\bibitem [{foo({\natexlab{a}})}]{footnote}%
  \BibitemOpen
  \bibinfo {note} {The relevant observable is $\braket{\Psi}{F_\gamma}$ where
  $\ket{\Psi}$ represents a floppy mode and $\ket{F_\gamma}$ are the forces on
  the network's nodes that arise due to the application of a strain $\gamma$.
  Since in mechanical equilibrium of our floppy networks
  $\ket{F_\gamma}\!=\!{\cal S}^T\ket{\partial r/\partial\gamma}$ (see
  Appendices for details), and floppy modes satisfy ${\cal S}\ket{\Psi}\!=\!0$,
  then it follows that $\braket{\Psi}{F_\gamma}\!=\!0$.}\BibitemShut {Stop}%
\end{thebibliography}
%

\end{document}